# On the Same Page

Dimensions of Perceived Shared Understanding in Human-AI Interaction

Qingyu Liang
Faculty of Business
City University of Macau
Taipa, Macau
qingyu.liang@outlook.com

Jaime Banks
School of Information Studies
Syracuse University
Syracuse, NY, USA
banks@syr.edu

## ABSTRACT

Shared understanding plays a key role in the effective communication in and performance of human-human interactions. With the increasingly common integration of AI into human contexts, the future of personal and workplace interactions will see a greater prevalence of human-AI interaction (HAII) in which the perception of shared understanding (PSU) will be important. Existing literature has addressed the processes and effects of PSU in human-human interactions, but the construal remains underexplored in HAII. To better understand PSU in that context, we conducted an online survey to collect user reflections on interactions with a large-language model when its understanding of a situation was thought to be similar or different from the participant's. Through inductive thematic analysis, we identified eight dimensions comprising PSU in human-AI interactions. The descriptive framework we derive supports an operational characterization of PSU and serves as a springboard for future work into the phenomenon.

## AUTHOR'S NOTE

A self-report scale based on this work has been developed and is now available here: Liang, Q., & Banks, J., (2025). Perceived shared understanding between humans and artificial Intelligence: Development and validation of a self-report scale. *Technology, Mind, and Behavior, 6*(1). https://doi.org/10.1037/tmb0000161

## 1 Introduction

With the advance of large language model (LLM) technologies, many AI systems are now capable of generating natural language to interact with humans, either through text-based or speech-based conversations. Additionally, through data-driven model training, some LLMs can learn from past interactions to refine responses over time, increasingly aligning with user intent and preference (Long et al., 2022). As a result, users often have the impression they are "talking to something intelligent" (Arcas, 2022, para. 9). These impressions can be formed across various contexts: AI is increasingly integrated into workplaces, where it serves as an efficient and reliable teammate in work-related tasks (e.g., Liang, 2024; Seeber et al., 2020). In the context of "hybrid intelligence" (Dellermann et al., 2019), humans and AI work together, leveraging their respective strengths to improve task performance and decision-making accuracy. Outside of the workplace, social chatbots have demonstrated the ability to engage in empathetic, humanlike conversations, offering companionship and emotional support. This capability has led to its adoption in various personal roles, such as AI pets (Melson et al., 2009), companions (Banks, 2024), friends (Brandtzaeg et al., 2022), and sex partners (Döring & Poeschl, 2019), fostering intimate relationships with humans. Extant literature indicates that actual and perceived understanding are key to these roles within human-human relationships, however they are not yet well understood in human-AI interactions. This work takes a step toward bridging that gap by inductively identifying dimensions of perceived shared understanding across varied human-LLM exchanges.

### 1.1 Shared Understanding, Briefly Explained

In considering shared understanding (SU), the concept of "sharedness" typically refers to "the degree to which members' mental models are consistent or converge with one another" (Mohammed et al., 2010, p. 880), which would lead to team members hold "common expectations for the task and team" (Cannon-Bowers et al., 1993, p. 236). Mental models are internalizations of knowledge about phenomena in the world—the networks of information humans or AI may hold about a situation or thing or idea (Craik, 1943; Mohammed et al., 2010). Through this lens, SU can be understood as the degree to which information in one agent's mental model matches the knowledge in another's, or not (see Cannon-Bower et al., 1993; Jonker et al., 2010). In other words, it is "the overlap of understanding and concepts among group members" (Mulder & Swaak, 2002, p. 36). Scholars have defined shared understanding in other ways as well. Smart et al. (2009, p. 1) describe it in terms of its outcomes as "the ability of multiple agents to coordinate their behaviors with respect to each other in order to support the realization of common goals or objectives." In a more nuanced fashion, Bittner and Leimeister (2014, p. 144) define it as "the degree to which people concur on the value of properties, the interpretation of concepts, and the mental models of cause and effect with respect to an object of understanding." In the face of these varied conceptualizations, we offer a synthesized definition that can be reasonably applied across both humans and AI: Shared understanding is an emergent state (Cash et al., 2020) of alignment between agents' knowledge bases (Kleinsmann et al., 2010), between the organizations of that knowledge (Hinds & Weisband, 2003), and between meanings ascribed to that knowledge (see Bohm, 1996).

When humans have a common understanding among themselves, it helps them better understand and interpret the information that is shared (Krauss & Bricker, 1990). According to McComb et al. (1999), people need to have a shared understanding of the path to take in order to finish a shared task before they can begin working together on it. Therefore, shared understanding influences members of a group to collaborate effectively and further influences collaborative performance (Mathieu et al., 2000). In the context of human-AI interaction, shared understanding becomes even more crucial as it bridges the gap between human expertise and AI's computational capabilities, ensuring smoother, more productive communication (Zhang et al., 2021). The degree to which AI can understand human expectations has significant implications for its reliability in tasks such as driving cars, diagnosing diseases, and performing other human-centric functions (Mitchell & Krakauer, 2022). In more social interactions, SU supports the quality of exchanges (Donnellon et al., 1986) between humans and AI, and it enables interactants to more accurately anticipate each other's needs and actions (Rico et al., 2008) and to realize "success of coalition operations" (Smart et al., 2009, p. 1). Past scholarship indicates that SU could contribute to improve collaboration performance (Mathieu et al., 2000), enhanced interactants satisfaction (Langan-Fox et al., 2004), and the reduction of iterative loops and redundancy (Kleinsmann et al., 2010). Conversely, when interactants lack a shared knowledge base, frameworks, or understanding of essential ideas, individuals probably will hold on different views on "what they should be doing and why" (see Ko et al., 2005, p. 64), it can impede effective interaction.

## 1.2 The Perception of Shared Understanding

Perceived shared understanding (PSU) is not a matter of manifest alignment of mental models. Rather, PSU comprises an individuals' *interpretation* of knowledge and meaning sharedness. Burtscher and Oostlander (2019) characterize the perception of mutual understanding as a team cognition phenomenon marked by "perceptions of the overall state of a team's knowledge structure, that is, perceptions of the extent to which a common knowledge structure exists" (p. 100). In human interactions, several key factors contribute to PSU, such as feeling understood when we talk, agreeing on what's important, having similar prior experiences, and having an agreed-upon way of getting work done (Cohen et al., 1999; Ko et al., 2005). The literature in this domain is quite limited with respect to artificial intelligence. Some work draws on models of PSU among humans and adapts them for AI in applied teaming, measuring them heuristically as matching for task completion strategy, communication, and information sharing (Schelble et al., 2024). Although AI can mimic people's conscious states to some extent, it is still different from people in many ways (see Grace et al., 2018; Korteling et al., 2021), and it is yet unclear whether and how people take ontological similarities and differences into account when considering PSU in HAII. Moreover, all work done in this place has unfolded in formal human-machine teaming contexts without consideration for how PSU may function in more general, everyday interactions with non-expert users. We argue, in line with Burtscher and Oostlander (2019), that humans' PSU is likely more socially important than whether the AI actually has such an understanding.

The limited scholarship in this domain highlights a key knowledge gap: It is yet unknown how people may interpret the nature of shared understanding between themselves and social-AI interlocutors. We do not focus here on whether AI truly possesses understanding abilities, rather on how individuals *interpret* an AI as demonstrating understanding that aligns with their own. This shift in emphasis moves the question from purely philosophical or technical one to a sociotechnical one (Liang, 2024; Makarius et al., 2020). Due to the differences between human and AI interactants—communication modalities, feedback patterns, emotional interaction, error handling, and ostensible cognitive capacities—we do not assume existing knowledge about PSU among humans is necessarily applicable to the emerging social-cognitive dynamics of HAII. Thus, this exploratory study aims to answer the following research question: ***What dimensions constitute humans' subjective perception of shared understanding with an AI?***

## 2 Method

To address the posed research question, we conducted an online survey soliciting people's real past experiences with an AI—specifically a large language model (LLM) accessed through a natural-language chat interface—and captured their evaluations of the AI's convergent or divergent understanding of the shared exchange. These elicited experiences were subjected to inductive thematic analysis to identify the prominent dimensions of perceptions of shared understanding. The study materials, anonymized data, and analysis documentation can be found in the online supplement for this project: https://osf.io/m9uwk/

### 2.1 Sampling and Recruitment

Participants were recruited in two phases and comprised convenience samples of LLM users aged 18 years and older. In the first recruitment phase, we solicited participants from the researchers' own networks and AI-relevant online forums and groups (e.g., LinkedIn, Facebook, Reddit); those participants were incentivized with a random drawing for a gift card. In the second recruitment phase, we invited undergraduate and graduate students from a university research participant pool; those participants received course credit for participation. In both cases, participation invitations were posted to relevant platforms and participants could click on the survey link in the invitation to be redirected to the online survey platform.

After collecting the questionnaire data (as described below), we cleaned the data of incomplete surveys and bot responses (made evident through attention checks and reviews of qualitative responses). This cleaning resulted in $N = 162$ (Phase 1: $N = 59$, Phase 2: $N = 103$) cases analyzed. Participants reporting successful interactions totaled 74, while those with unsuccessful interactions amounted to 88. The sample comprised 72 women, 85 men, one transwoman, one gender-nonconforming individual, and three not identifying; participants were White/Caucasian ($n = 46$), Black/African American (40), Asian (58), Hispanic (3), mixed-race or multi-ethnic (3), Indian (1), with 8 not identifying. The mean age was 26.17 years ($SD = 7.27$, range 18-62). Participants' educational backgrounds included 74 with a bachelor's degree, 47 with a master's degree, 28 with high school education, 11 with a doctoral degree, and three who did not disclose their education. Participants came from a variety of nationalities, with most from the United States (67), followed by China (34), and Nigeria (24). Other nationalities included India (11), Korea (3), Germany (2), and one participant each from Albania, Australia, Canada, Dominican Republic, the Netherlands, Finland, Ghana, Indonesia, Iran, Israel, Mongolia, Serbia, Turkey, and Portugal. Seven participants chose not to disclose their nationality.

## 2.2 Procedures and Elicitations

The survey consisted of three sections. First, participants were asked to share either a successful or unsuccessful interaction (randomly assigned) with an LLM by copy-pasting up to three conversational turns into an open-response field in the survey. Using the notions of "successful" and "unsuccessful" interactions as anchors for interaction selection ensured a variety of conversation quality (in addition to varied subject matter), since shared understandings or shared mental models are generally understood to produce more successful interactions. Subsequently, participants answered three open-ended questions eliciting perceptions of (un)shared understanding. Lastly, we gathered demographic information from participants, including age, gender, racial or ethnic background, highest level of education, and nationality (captured through open-response fields and then coded).

The focal LLM for this investigation was ChatGPT. We chose ChatGPT because the LLM was first to be mainstreamed, is widely used across professional and personal contexts, and laid the groundwork for many people's understandings of what social AI is and what it is capable of. It has also been the subject of debates as to whether or not it can meaningfully understand the world through its exchanges or is simply fluent in human languages. This broad adoption, varied use, and varied interpretation of capacities affords consideration of a range of users and use cases.

Among the three open-ended elicitations, two comprise the primary data analyzed here. To draw out participants' subjective perceptions of how ChatGPT's understanding and their own understanding of the situation did (or did not) align, participants were asked to think about their selected interaction with ChatGPT and then give open-ended responses to: a) How did you and the AI seem to be understanding the situation similarly? b) How did you and the AI seem to be understanding the situation differently? Attending to both convergent and divergent understandings was intended to generate data that would indicate dimensions of PSU at both high and low levels.

## 2.3 Analytical Approach

We conducted a thematic analysis following the analytical phases described by Braun and Clarke (2006) for analyzing and reporting various dimensions of people's subjective experience within the survey data. We began by thoroughly reading the survey data to become familiar with its content. Next, the first-listed author coded that data by assigning codes to data units (unitized as complete thoughts). Data was coded separately in two sets—one set for responses to the question about similar understanding and one set for the question about different understanding. Of note, responses across the two conditions were combined for each question, since the conditions were intended only to ensure qualitative variation in responses. These codes were then iteratively reduced into categories and then compared across the two questions asked of all participants (i.e., how was the AI seeming to understand similarly and differently) to collapse them into higher-order themes. The second-listed author conducted a face-validity check and identified "devil's advocate" opportunities to reconfigure the thematic hierarchy. Having identified some overlap in initial themes, both authors discussed and refined these themes, ensuring internal coherence and resolving the devil's-advocate challenges. See the online supplements for complete data, initial codes, code reduction, and devil's advocate challenges

## 3 Results

We induced eight key themes in PSU between perceiving humans and an interacting AI. Below, we describe five themes inherent to similar understanding and eight themes inherent to divergent understanding. Some dimensions are mentioned in both categories yet are expressed in positive and negative valences—that is, in some form, "understanding X similarly" and "understanding X differently." When this occurred, we integrated contrasting descriptions under the positively valenced theme. The labels "S" and "D" have been introduced to differentiate between responses to the question about similar (i.e., shared) understanding ("S") and dissimilar understanding ("D"), providing a clear framework for analyzing the distinct patterns within each condition. We summarize the themes, sub-themes, and their (dis)similar origins in Table 1, and unpack them in the following pages.

## 3.1 Fluency

Fluency is a matter of mutual understanding of words exchanged between humans and AI. When people use words, whether text-based or spoken, PSU can consist of seeing AI as able to accurately decode and interpret their meaning. At the same time, PSU also requires humans' understanding the meaning of an AI's words, such that exchanged messages are encoded and decoded correctly between humans and AI. This theme consists of three key aspects: Input comprehension, intention comprehension, and aligned interpretations. Input comprehension emphasizes whether AI can understand user prompts, questions, instructions, or requests. For example, one participant explained, "it didn't understand my question correctly" (D43). Another participant observed that "I can use industry speak, and it understands what I mean" (S103). Some individuals noted, "I think ChatGPT captures some keywords in my prompts" (S47) and "AI knows the words I used even acronyms" (S103). Intention comprehension means that an AI not only accurately decodes the user's input but also appears to grasp the expectations, goals, and purposes conveyed through their words. For instance, one participant described a moment of didn't shared clarity: "AI didn't know the size and design I wanted" (D52), while another participant felt the AI successfully interpreted their request, stating, "It also looks at the entire prompt as a bigger picture" (S46).

## 3.2 Aligned Operation

Participants sometimes emphasized a perception that they and the AI achieve alignment during interactions, operating with a similar mindset in terms of values and approaches, ensuring that both parties work toward the same objectives with a common framework. The aligned operation across three key areas: Aligned perspectives, aligned approaches, and aligned goals. Aligned perspectives refer to whether AI and users share the same viewpoint or interpret situations through a similar lens. Participants reported observing moments when their perspectives are different from the AI's; some people hold the view that their unique personal experiences and emotions often differ from AI's interpretations, making alignment challenging. For instance, one participant explained, "I think I have a different... perspective than it" does (D145). Another indicated, "My unique perspective, emotional understanding, and personal context differ from those of the AI" (D147). Aligned approach refers to how well the AI's methods correspond with those of the user. When AI strategies mirror those of the user, it appears to foster a sense of collaboration and alignment. One participant remarked, "We both tried to find the optimal price that reaches profit at its maximum level" (S33). Another participant noted, "It was able to approach the problem in the same way that I would've approached it" (S127). Aligned goals seem evident to users when AI align with their intended outcomes or objectives. This means that the AI not only recognizes what the user is trying to achieve (see fluency discussed above) but also operates in a way that drives the interaction toward a common goal. As one participant described, "I believe we were heading in the same direction" (D119), while

**Table 1**. Overview of Themes In Perceived Shared Understanding in HAI

| Themes | Explanation | Subdimensions | Sim. | Diss. |
|---|---|---|---|---|
| Fluency | AI does (not) grasp human input and intentions accurately, resulting in a proper words exchange. | • Input comprehension<br>• Intention comprehension | √ | √ |
| Aligned Operation | AI (dis)aligns with humans in values and actions to ensure consistency in working toward shared objectives. | • Aligned perspectives<br>• Aligned approaches<br>• Aligned goals | √ | √ |
| Fluidity | AI can (not) successfully engage in a fluid, back-and-forth information exchange with user. | • Continuity<br>• Adaptation<br>• Consistency | √ | √ |
| Satisfaction with Outcomes | User is (dis)satisfied with the AI's outcomes. | • Alignment with user expectations<br>• Accuracy of answers<br>• Task execution capabilities | √ | √ |
| Contextual Awareness | AI does (not) understand or adapt to the interaction context. | • Contextual understanding<br>• Context-matched responses<br>• Alignment with reality | √ | √ |
| Lacking Human-like Abilities | AI lacks traits that resemble human interaction. | • Proactive engagement<br>• Subjective insights | | √ |
| Computational Limitation | AI has inherent technical constraints that restrict its ability to perform certain tasks. | • Computational limitations | | √ |
| Suspicion | User doubts AI's answers. | • Answer reliability<br>• Answer familiarity<br>• Answer consistency | | √ |

*Note: Sim. = themes derived from explanations of perceived similar understanding. Diss. = themes derived from explanations of perceived dissimilar understanding.*

another more overtly stated, "We have similar understanding in terms of task goal" (S33).

### 3.3 Fluidity

This theme considers the perceived ease of interaction between users and AI, emphasizing how fluid, continuous exchanges are maintained through conversation consistency, adaptation to user input, and the ability to respond naturally. Conversation continuity refers to whether AI and users can maintain a coherent and uninterrupted dialogue. Participants noted instances where the AI failed to follow the conversation when it forgot users' previous inputs (e.g., questions) or suddenly shifted to unrelated content. For example, one participant expressed frustration: "AI seemed to have forgotten that it was fantasy" in the middle of a fantasy narrative (D109). Another emphasized the need for "continuity. Our conversations are always centered around a theme" (S86). Conversation adaptation involves the ability of both users and AI to adjust their communication styles in response to one another, leading to more relevant and accurate turns. Participants indicated the AI could often adapt based on their inputs, generating more relevant responses as they refined their instructions. One user explained, "I rewrote the question in a more descriptive way so that it made sense to the AI and me" (S5). Finally, conversation consistency refers to the AI's ability to maintain logical consistency throughout the interaction. Users expressed concerns about inconsistencies in the AI's responses, with one participant noting, "I was hoping AI would have some embedded commitment to logical consistency in its utterances" (D89). Another user noted that: "Both understood the question was related to [the topic]. However, the AI missed the mark on most of what I was trying to pin down" (S88).

### 3.4 Outcome Satisfaction

Users sometimes characterized PSU through positive evaluations of the AI's performance, focusing on three main components: Alignment with users' expectations, accuracy of answers, and task execution capabilities. The first component is whether AI's responses meet users' expectations, including its ability to offer helpful feedback, satisfy their needs, or provide answers that leave them feeling positive. For example, one participant remarked, "AI is responding perfectly with what was expected" (S129). However, when AI failed to meet these expectations, users were left dissatisfied, as another participant noted, "what it gave was out of place for me and the [things] I intended to embark on" (D71). The accuracy of answers highlights instances where the AI provided incorrect or irrelevant responses. For example, one user reported that "the AI was just repeating the wrong answers" (D69), while another noted that "ChatGPT wrongly assigned answers that did not help" (D84). Task-execution capabilities reflect its ability to complete specific tasks as instructed by the user. Participants appeared to appreciate when AI successfully carried out their requests, as highlighted by one user who said, "I wish AI can give me a comprehensive analysis, and obviously it did" (S35).

### 3.5 Contextual Awareness

Responses interpreted as perceiving contextual awareness reflected an AI's ability to comprehend the situational background behind users' inputs and to generate responses that are appropriately matched to that context. It consists of three key aspects: Contextual understanding, context-matched responses, and alignment with reality. Contextual understanding refers to AI's perceived capacity to grasp the background or implied information within a user's input, even if not explicitly stated. For instance, one participant indicated an AI can't share understanding because AI "fails to answer the questions implied by the questions, but only answers them literally" (D110). Another participant noted they and the AI "Both understood the question was related to music related to Casablanca" (S88) On the other hand, context-matched responses focus on the degree to which AI's generated answers are well-matched to the situation. One participant mentioned, "it would not give me a generic response" (S143). Another participant indicated the AI would "generate text based on the input it receives. This enables it to provide relevant and contextually appropriate responses, making it appear as though we understand the situation similarly" (S29). Alignment with reality refers to whether the AI's suggestions and responses are realistic and relevant to current real-world conditions. For example, user may perceive AI can't share understanding when "AI understood the situation but not up to date with the current tax law" (D77), or "The road trip suggestion (provided by AI) was a bit unrealistic" (D120).

### 3.6 Lacking Humanlike Abilities

This theme, exclusive to those in the unsuccessful-interaction condition, suggests AI has low PSU when it falls short of replicating human interaction impacts, particularly in terms of proactive engagement and the ability to have human-like subjective insights. First, AI may fail to proactively engage the user, not taking initiative in conversations by seeking clarification or ensuring user satisfaction. Unlike humans, who naturally seek feedback to refine understanding (see Flavell, 1979), after generating the response the AI would not further verify if the answer fully meets the user's needs. Meanwhile, AI always provides feedback only after receiving input from people and does not proactively ask for information from them. As one participant noted, "it didn't seek my thoughts or opinions about whether it answered my question fully" (D61). Another said, "I need to ask questions, in order to give specific answers I want to see. But GPT won't ask." (D66). Second, some individuals consider that AI lacks the capacity for subjective thinking, offering only objective, fact-based responses without the personal insights or opinions that human interactions often rely on. As one participant observed, "My understanding is more related to my personal life experiences and academic knowledge, while AI's is based on information from online research" (D79).

### 3.7 Computational Limitations

This theme focuses on the functional limitations of AI, which restrict its ability to perform certain tasks effectively; in the context of the elicitation, we interpret this as being tied to reduced PSU. As a technology, AI has inherent constraints that prevent it from completing specific tasks, such as accessing real-time information. For example, participants noted that "it does not have real-time access to current events or updates beyond its last training data" (D29) and that "it could not find the article and was confused on how to find it even though I gave the full title" (D11). These functional shortcomings appear to impact AI's ability to handle users' some specific tasks, leading users to perceive that the AI cannot share a true understanding of their needs due to it those limitations.

### 3.8 Suspicion

This theme captures participants' doubts regarding the outputs generated by AI, focusing on three key aspects: Answer reliability, answer familiarity, and answer consistency. Notably, we distinguish this suspicion from the prior, satisfaction-focused themes in that suspicion does not necessarily mean (dis)satisfaction but is instead the user's own evaluative processes by which dissatisfaction could emerge. Here, answer reliability refers to users' concerns about the data reliability and the logic of the AI's responses. For instance, one participant remarked, "The logic of the AI seems confused..." (D61). Another noted, "It is better in terms of accuracy in terms of specifics, but weaker in terms of data validity" (D110), highlighting concerns about the soundness of the AI's responses. Answer familiarity relates to the AI providing responses that seem unusual to the user. People expressed feeling suspicious when presented with unfamiliar data, leading to reluctance in accepting the AI's suggestions. For example, one participant stated, "The AI stated so many other books of account that I have never heard of, which made me suspicious, and I only took those that are familiar to me" (D117). Answer consistency focuses on discrepancies between the AI's output and users' prior knowledge or other reference materials. Participants expressed concern when the AI's responses contradicted information they had learned elsewhere. One participant noted, "The AI gave a different answer to what the book is entirely about" (D111), while another mentioned, "I have never heard of them, and that made me suspicious" (D117).

## 4 Discussion

Motivated by a lack of data-driven research into the human perception of shared understanding (PSU) between interacting humans and AI, we conducted an inductive analysis of LLM users' reflections on instances when an LLM was thought to exhibit (dis)similar understanding in past interactions. Through that analysis, we identified eight dimensions of PSU across (un)successful interactions. We offer a thematic hierarchy describing how PSU manifests as a function of interaction fluency, operation, fluidity, outcome satisfaction, contextual understanding—and additionally for a lack of shared understanding—the absence of human-like abilities, computational limitations, and user suspicion.

Our findings suggest that PSU in HAII shares significant overlap with human PSU constructs such as fluency, operational alignment, outcome focus, interaction fluidity, and suspicion (i.e., trust-related considerations). However, other dimensions are likely exclusive to PSU for artificial interlocutors: Contextual awareness concerns, lack of human-like abilities, and computational limitations. Although the overlap with dimensions known to human PSU could not be assumed, it is not surprising since AI is increasingly designed to mimic human behaviors across a range of tasks traditionally performed by humans. That is, because the performance of interaction delivers, by design, many of the same cues as human interactants, it is reasonable that those cues could trigger some of the same interpretations of the machine's understanding of a situation; the construal of understanding (dis)alignment is shaped by machine's humanlike sociality (Fong et al., 2003). The dimensions that diverge from known human-PSU considerations are interpreted to suggest that an agent's ontological category is highly salient to the perceiving human; that cognitively efficient categorization is likely anchoring the evaluation of SU. We interpret these patterns in PSU by considering them in relation to an extant model of manifest shared understanding, before extending that model for additional dimensions inherent to PSU. The model of shared understanding presented by Gomes and colleagues (2017) characterizes SU among humans not as a state or condition, but instead as a communicative process—ideas are shared, meaning is determined through negotiation, often that negotiation unfolds through mediating artifacts that are differently engaged at agent boundaries (cf. Star & Griesemer, 1989); determined meaning is then engaged to motivate collective or cooperative action. Central to that model are interactions and meaning-making, aligning with two classes of our observed themes (interaction and evaluation). We also observed a third class of PSU dimensions that permits extension of that existing model (Figure 1) to account for additional construals related to the *kind* of thing an interacting AI is thought to be. Altogether, we argue that PSU is an inherently communicative phenomenon in which the construal may be one of the interaction, the outcome and its value, and/or of one's beliefs about the agent and the shared context.

### 4.1 Potential Dimensionality

We infer from this analysis a potential clustering of PSU into three dimensions of PSU, guided by extant scholarship.

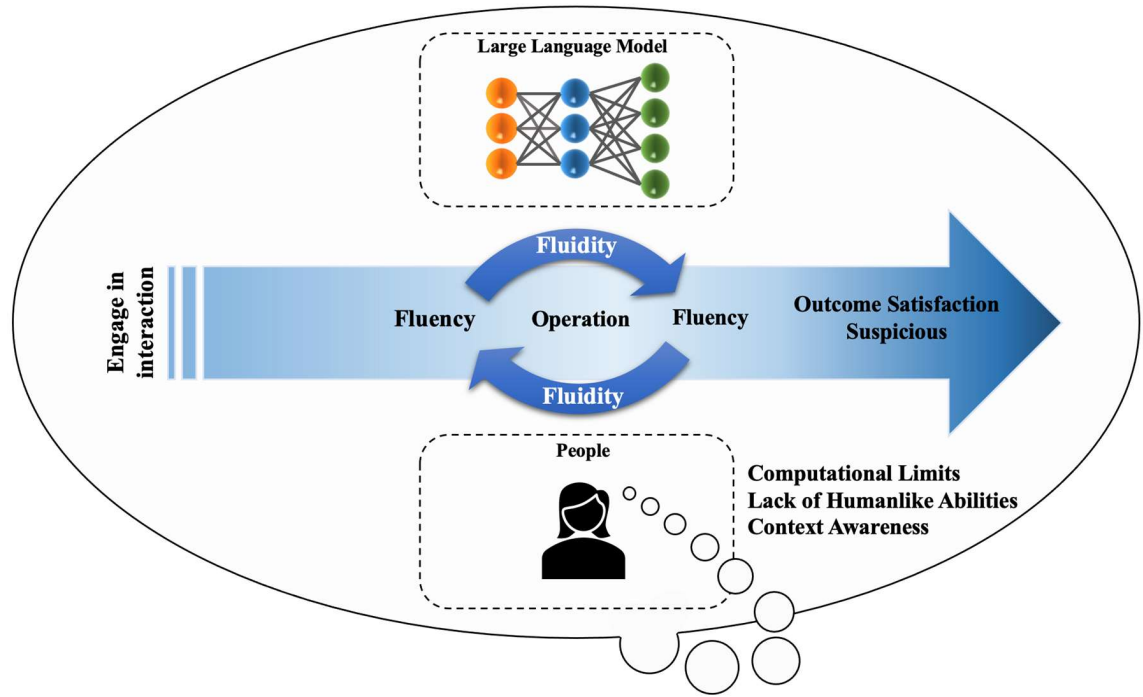

**Figure 1: PSU dimensions for human-AI interaction, superimposed on Gomes et al.'s (2017) model for manifest shared understanding.**

Fluency, fluidity, and operational alignment are all PSU dimensions that reflect the human assessment of **interaction processes**. Specifically, an interaction may be characterized by effective encoding and decoding of messages facilitating meaningful exchanges of information between humans and AI, enabling both parties to better understand each other's expectations and requirements (Clevenger, 1971; Roberts et al., 1987). We observed that conversational flow and easy exchange were central to PSU; those factors are also known to contribute to individuals' sense of social unity, effectively "constructing a shared sense of 'us'" (Koudenburg et al., 2014, p. 363). Finally, operational alignment, or the correspondence between the actions and goals of both actors, is known to reinforce a sense of connectedness between individuals during interactions through behavioral coordination (Miles et al., 2010; Schmidt & Richardson, 2008).

PSU may also comprise **evaluative dimensions** related to the quality, validity, or utility of the interaction output, such that PSU is not only part of the process of interaction but continues to be interpreted after the formal interaction is complete. In this study, participants indicated outcome (dis)satisfaction, focusing on the accuracy and appropriateness of the information provided during the interaction (known among humans; Farace et al., 1978), perceived functionality of the AI since users are more likely to rely on systems that consistently execute tasks competently (Lee & See, 2014), and alignment with expectations as satisfaction arises when outcomes meet or exceed expectations (Oliver, 1980; Parasuraman & Riley, 1997). Additionally, some responses revealed suspicion toward the AI's outputs, an indication of compromised trust, which aligns with Ivarsson and Lindwall's (2023) findings for humans: "When suspicion is roused, shared understanding is disrupted" (p. 1).

In addition to perceptions related to the actual interaction and its outcome, participants also demonstrated a set of higher-order **ontological construals** that would likely be taken for granted in human-human PSU. Specifically, PSU for AI may rely on considerations of the AI itself (computational limitations and a lack of human-like abilities) or of the relationship between the AI's perceived (non)situatedness in a shared context. It is not entirely clear from our data whether the human-like limits and performed/limited contextual awareness were a function of comparing the AI to oneself or a function of comparing the AI to one's assumptions about how other humans function and what they can do (e.g., Edwards, 2018). The latter interpretation would semantically align with many points in our data and with notions of a general theory of mind that people may hold for other agents' internal states (both human and machine; Wang et al., 2021) and with holding mental models for what agents in different ontological categories can do (Banks, 2020).

### 4.2 Limitations and Future Directions

Our study primarily focuses on humans' perception of shared understanding with LLM-based, general-purpose AI. However, we acknowledge the diversity of AI and human-AI interaction modes, including AI embedded in physical robots, intelligent systems, and other interaction formats such as voice and multimodal communication. These different forms may introduce distinct factors inherent to perceived shared understanding (e.g., the introduction of dialect or embodied abilities). Therefore, future research could extend our investigation to other AI types and interaction modes, encompassing a larger sample size to enhance the generalizability of the findings.

Interpretive analysis of qualitative data is inherently subjective such that replication of this work by other interpreters would be productive. For the present study, we engaged best practices in face-validity checks of the first-author's coding, by the second author, and engaged in discussions to resolve any discrepancies, bolstering our confidence in the findings; the first author is a scholar of HAII in work contexts and the second in social contexts, ensuring that both applied and autotelic considerations would be made in data interpretation. We purposefully and systematically considered by positive and negatively valenced interactions across different populations and a range of topics/applications to ensure varied potentials; analysis was conducted through a data-up (rather than theory-down) approach to ensure openness to emergent patterns. Although we made some efforts to ensure the results' reliability, repeated validation of our proposed factors among additional, varied participants is necessary in the future.

Importantly, we have identified PSU dimensions that both align with and diverge from known dimensions in human interactions; these patterns present a range of future research possibilities. For instance, when humans hold PSU with AI, they may develop a sense of team spirit or form subjectively intimate connections (both themes previously discussed in human teams; see Kniel & Comi, 2021). In another vein, it could be that advances in actual and perceived fluidity/fluency improve the potential for AI to facilitate human self-efficacy in conversation in was that promote the experience of rewarding conversations (e.g., Rubin et al., 1993). In order to systematically test these and other potentials, though, appropriate measurements for the identified PSU dimensions should first be developed and validated (see Liang & Banks, 2025). Most broadly, considering PSU as an emerging process rather than static or cross-sectional state could help to inform understandings of how PSU could matter in long-term HAII (such as in companionship scenarios or co-working contexts) to influence a range of outcomes—from the efficiency and correctness of task-focused interactions to the depth and gratifications of social interactions (e.g., Machia et al., 2024).

Dimensions of Perceived Shared Understanding in HAII